\documentclass[12pt]{iopart}

\begin{document}

\title[ ]{Scalar-tensor theory with enhanced gravitational effects. }
\author{F. O. Minotti}

\address{Departamento de F\'{\i}sica, Facultad de Ciencias Exactas y Naturales,
Universidad de Buenos Aires - INFIP (CONICET) }

\begin{abstract}
It is shown that a Brans-Dicke scalar-tensor gravitational theory, which
also includes Bekenstein's kind of interaction between the Maxwell and
scalar fields, has a particular kind of solutions with highly enhanced
gravitational effects as compared with General Relativity, prone to
laboratory tests.
\end{abstract}

\maketitle

\section{Introduction}

Scalar-tensor (ST) gravitational theories are attractive candidates for
extensions of General Relativity (GR). Part of their appeal comes from the
fact that they are induced naturally in the reduction to four dimensions of
string and Kaluza-Klein models\cite{fujii_book}, resulting mostly in the
form of a Brans-Dicke (BD) type of ST theory\cite{brans_dicke}, often
involving also non-minimal coupling to matter, leading to weak equivalence
principle violations. Also, ST theories are possibly the simplest extension
of GR that could accommodate cosmological issues as inflation and
universe-expansion acceleration, as well as possible space-time variation of
fundamental constants\cite{bekenstein}. On the other hand, observational and
experimental evidence puts strong limits to the observable effects of a
possible scalar field. For example, in the case of a massless scalar the BD
parameter $\omega $ is constrained by careful Solar-System experiments to be
a large number ($\omega >5\times 10^{5}$)\cite{bertotti}. This has not been
favorable for the ST theories, although it was found that some Kaluza-Klein
models allow ST theories with arbitrarily high values of $\omega $\cite
{chavineau}, and a "least-coupling principle" was proposed in string
theories to the same effect\cite{damour_polyakov}. Closely related to the ST
theories is the question of the space-time variation of fundamental
constants. The proposal of models compatible with all known constraints are
basically of the Bekenstein's type\cite{bekenstein}, involving the coupling of
the electromagnetic field to a scalar field\cite{sandvik},\cite{damour}. The
resulting violations of the equivalence principle are still below the actual
experimental limits, either in the propagation of light\cite{fujii}, or in E
\"{o}tv\"{o}s experiments. The purpose of the present work is to show that a
ST theory of the BD type, that also includes a direct interaction between
the Maxwell and the scalar fields, as in Bekenstein's theory, has a special
kind of solutions where the metric effect of the electromagnetic field is
strongly enhanced, as compared to the case of GR. This enhanced effect could
be observed with relative ease in laboratory experiments with present
technology.

\section{ST\ theory}

We will consider a ST theory of the BD type with inclusion of a Bekenstein's
direct interaction of scalar and Maxwell fields, with action given by (SI
units are used)%
\begin{eqnarray}
S &=&\frac{c^{3}}{16\pi G_{0}}\int \sqrt{-g}\phi Rd\Omega -\frac{c^{3}}{16\pi G_{0}}\int \sqrt{-g}\frac{\omega \left( \phi \right) }{\phi }\nabla
^{l}\phi \nabla _{l}\phi d\Omega  \nonumber \\
&&-\frac{\varepsilon _{0}c}{4}\int \sqrt{-g}\lambda \left( \phi \right)
F_{lm}F^{lm}d\Omega -\frac{1}{c}\int \sqrt{-g}j^{l}A_{l}d\Omega -S_{mat}.
\label{SKK}
\end{eqnarray}%
In order to have a non-dimensional scalar field $\phi $ of values around
unity, in expression (\ref{SKK}) the constant $G_{0}$ representing Newton
gravitational constant is included, $c$ is the velocity of light in vacuum,
and $\varepsilon _{0}$ is the vacuum permittivity. $S_{mat}$ is the
non-electromagnetic action of matter. The other symbols are also
conventional, $R$ is the Ricci scalar, and $g$ the determinant of the metric
tensor $g_{lm}$. The Brans-Dicke parameters $\omega \left( \phi \right) $ is
considered a function of $\phi $, as it usually results so in the reduction
to four dimensions of multidimensional theories. The function $\lambda
\left( \phi \right) $ in the term of the action of the electromagnetic field
is of the type appearing in Bekenstein's theory. The electromagnetic tensor
is $F_{lm}=\nabla _{l}A_{m}-\nabla _{m}A_{l}$, \ given in terms of the
electromagnetic quadri-vector $A_{m}$, with sources given by the
quadri-current $j^{l}$.

Variation of (\ref{SKK}) with respect to $g^{lm}$ results in ($T_{lm}^{EM}$
is the usual electromagnetic energy tensor)%
\begin{eqnarray}
\phi \left( R_{lm}-\frac{1}{2}Rg_{lm}\right) &=&\frac{8\pi G_{0}}{c^{4}}%
\left[ \lambda \left( \phi \right) T_{lm}^{EM}+T_{lm}^{mat}\right] -\nabla
^{k}\nabla _{k}\phi g_{lm}+\nabla _{l}\nabla _{m}\phi  \nonumber \\
&&+\frac{\omega \left( \phi \right) }{\phi }\left( \nabla _{l}\phi \nabla
_{m}\phi -\frac{1}{2}\nabla ^{k}\nabla _{k}\phi g_{lm}\right) .  \label{Glm}
\end{eqnarray}

Variation with respect to $\phi $ gives
\[
\phi R+\left( \frac{d\omega }{d\phi }-\frac{\omega }{\phi }\right) \nabla
^{l}\phi \nabla _{l}\phi +2\omega \nabla ^{l}\nabla _{l}\phi =\frac{4\pi
G_{0}\varepsilon _{0}}{c^{2}}\phi \frac{d\lambda }{d\phi }F_{lm}F^{lm}, 
\]%
which can be rewritten, using the contraction of (\ref{Glm}) with $g^{lm}$
to replace $R$, as
\begin{equation}
\left( 2\omega +3\right) \nabla ^{l}\nabla _{l}\phi +\frac{d\omega }{d\phi }
\nabla ^{l}\phi \nabla _{l}\phi =\frac{4\pi G_{0}\varepsilon _{0}}{c^{2}}
\phi \frac{d\lambda }{d\phi }F_{lm}F^{lm}+\frac{8\pi G_{0}}{c^{4}}T^{mat},
\label{phi}
\end{equation}%
where it was used that $T^{EM}=T_{lm}^{EM}g^{lm}=0$.

Finally, the non-homogeneous Maxwell equations are obtained by varying (\ref
{SKK}) with respect to $A_{m}$,%
\begin{equation}
\nabla_{l}\left[ \lambda\left( \phi\right) F^{lm}\right] =\mu_{0}j^{m}.
\label{Maxwell}
\end{equation}
with $\mu_{0}$ the vacuum permeability.

Having included $G_{0}$, it is understood that $\phi$ takes values around $
\phi_{0}=1$. Eqs. (\ref{Glm})-(\ref{Maxwell}) with $\lambda\left(
\phi\right) =1$ represent the original Brans-Dicke system if $\omega$ is
taken as constant.

\section{\protect\bigskip Special solutions}

A rapid inspection of Eqs. (\ref{Glm}) and (\ref{phi}) shows that
corrections to the metric generated by $T_{lm}^{mat}$ alone are in general
quantified by $\delta R\sim G_{0}\varepsilon_{0}F_{lm}F^{lm}/c^{2}$, as in
GR. However, for those solutions satisfying $\nabla^{l}\nabla_{l}\phi=0$,
one has instead $\delta R\sim\sqrt{G_{0}\varepsilon_{0}F_{lm}F^{lm}\left(
d\lambda /d\phi\right) /\left( c^{2}d\omega/d\phi\right) }/L$, with $L$ a
characteristic length-scale of variation of the fields. In this way,
depending on how precisely the condition $\nabla^{l}\nabla_{l}\phi=0$ holds,
and on the possible value of $\left( d\lambda/d\phi\right) /\left( d\omega
/d\phi\right) $, the system of equations considered allows solutions with
enhanced gravitational effects. We now analyze some cases where this kind of
possible solutions of the system (\ref{Glm})-(\ref{Maxwell}) exist, in the
weak field limit of the equations.

In the weak field approximation, for variations of $g_{lm}$ around the
values $\eta _{lm}$ taken as those of flat Minkowsky space with signature
(1,-1,-1,-1), so that $g_{lm}=\eta _{lm}+h_{lm}$, and of $\phi $ around $
\phi _{0}=1$, so that $\phi =1+\sigma $, we have
\[
R_{lm}-\frac{1}{2}R\eta _{lm}=\frac{1}{2}\left( -\eta ^{ik}\partial _{ik}
\overline{h}_{lm}+\partial _{il}\overline{h}_{m}^{i}+\partial _{im}\overline{
h}_{l}^{i}-\eta _{lm}\partial _{ik}\overline{h}^{ik}\right) , 
\]%
with%
\[
\overline{h}_{lm}\equiv h_{lm}-\frac{1}{2}h\eta _{lm}, 
\]%
where%
\[
h\equiv \eta ^{ik}h_{ik}=-\eta ^{ik}\overline{h}_{ik}. 
\]%
The system (\ref{Glm})-(\ref{Maxwell}) can be written to lowest order in the
perturbations satisfying $\nabla ^{l}\nabla _{l}\phi =0$ (which should hold
at least up to order two in the perturbed fields), 
\begin{equation}
-\eta ^{ik}\partial _{ik}\overline{h}_{lm}+\partial _{il}\overline{h}
_{m}^{i}+\partial _{im}\overline{h}_{l}^{i}-\eta _{lm}\partial _{ik}
\overline{h}^{ik}=2\partial _{lm}\sigma ,  \label{Rlin}
\end{equation}
\begin{equation}
\left. \frac{d\omega }{d\phi }\right\vert _{\phi _{0}}\partial ^{l}\sigma
\partial _{l}\sigma =\frac{4\pi G_{0}\varepsilon _{0}}{c^{2}}\left. \frac{
d\lambda }{d\phi }\right\vert _{\phi _{0}}F_{lm}F^{lm},  \label{philin}
\end{equation}
\begin{equation}
\partial _{m}F^{lm}=\mu _{0}j^{m}.  \label{maxlin}
\end{equation}
While the equation $\nabla ^{l}\nabla _{l}\phi =0$ is written up to second
order as%
\begin{equation}
\partial _{ik}\sigma \left( \eta ^{ik}+\overline{h}^{ik}+\frac{\eta ^{ik}h}{2%
}\right) +\partial _{k}\sigma \left( \partial _{i}\overline{h}^{ik}+\eta
^{ik}\partial _{i}h\right) =0.  \label{phill}
\end{equation}%
The raising and lowering of indices are effected by the tensors $\eta ^{ik}$
and $\eta _{ik}$\ respectively. Even this much simpler system is rather
complex. From a practical point of view it is convenient to restrict the
choice of coordinates by requiring the additional four conditions on $
\overline{h}_{lm}$%
\begin{equation}
\partial _{i}\overline{h}^{ik}+\eta ^{ik}\partial _{i}h=-\frac{\partial
_{lm}\sigma \overline{h}^{lm}}{\partial ^{l}\sigma \partial _{l}\sigma }
\partial ^{k}\sigma ,  \label{gauge}
\end{equation}%
so that Eq. (\ref{phill}) reduces to%
\begin{equation}
\eta ^{ik}\partial _{ik}\sigma =0.  \label{dalamb}
\end{equation}

The equations (\ref{Rlin})-(\ref{maxlin}), together with (\ref{gauge}) and (
\ref{dalamb}), constitute an ovedetermined set of equations which has
solutions only for particular cases. As it stands, the system can be solved
independently for the electromagnetic field given the sources $j^{m}$ in (
\ref{maxlin}), and then determine $\phi$ using (\ref{philin}) and (\ref
{dalamb}), to finally obtain $\overline{h}_{lm}$ from Eqs. (\ref{Rlin}) and (
\ref{gauge}).

\bigskip In terms of the modulus of the electric and magnetic vector fields, 
$E$ and $B$, respectively, one has 
\[
F_{lm}F^{lm}=2\left( B^{2}-E^{2}/c^{2}\right) , 
\]
so it is immediately seen from Eq. (\ref{philin}) that a possible solution
exists in the case of static fields outside their sources. In effect,
consider for instance the case of a pure electrostatic field, so that, using
greek indices for the three spatial coordinates, one has for the
electrostatic potential $\varphi ,$ 
\[
E_{\alpha }=-\partial _{\alpha }\varphi , 
\]%
with%
\[
\partial _{\alpha \alpha }\varphi =0, 
\]%
outside the electric sources. In this static case Eq. (\ref{philin}) is
written as%
\[
\partial _{\alpha }\sigma \partial _{\alpha }\sigma =\frac{8\pi
G_{0}\varepsilon _{0}\left( d\lambda /d\omega \right) _{\phi _{0}}}{c^{4}}
\partial _{\alpha }\varphi \partial _{\alpha }\varphi , 
\]%
where we have written $\left( d\lambda /d\omega \right) _{\phi _{0}}\equiv
\left( d\lambda /d\phi \right) _{\phi _{0}}/\left( d\omega /d\phi \right)
_{\phi _{0}}$. In this way, solutions also satisfying Eq. (\ref{dalamb})
then exist if $\left( d\lambda /d\omega \right) _{\phi _{0}}>0$, given by
\begin{equation}
\partial _{\alpha }\sigma =\pm K\sqrt{\frac{8\pi G_{0}\varepsilon _{0}\left(
d\lambda /d\omega \right) _{\phi _{0}}}{c^{4}}}\partial _{\alpha }\varphi ,
\label{dsigstat}
\end{equation}%
where%
\begin{equation}
K\equiv \frac{\sqrt{8\pi G_{0}\varepsilon _{0}\left( d\lambda /d\omega
\right) _{\phi _{0}}}}{c^{2}}.  \label{Kconst}
\end{equation}

If $\left( d\lambda /d\omega \right) _{\phi _{0}}<0$ one has solutions
instead only for a magnetostatic field outside its sources, so that $
B_{\alpha }=\partial _{\alpha }\psi $, with $\partial _{\alpha \alpha }\psi
=0$, given by%
\[
\partial _{\alpha }\sigma =\pm cK\partial _{\alpha }\psi . 
\]

For clarity sake we can assume that $\left( d\lambda/d\omega\right)
_{\phi_{0}}>0$, as all derivations are totally analogous in the case $\left(
d\lambda/d\omega\right) _{\phi_{0}}<0$, with $c\psi$ taking the place of $
\varphi$.

Gravity effects can be quantified in the weak field approximation as a
three-dimensional force per unit mass, given in the static case considered by%
\begin{equation}
f_{\beta}=-\frac{c^{2}}{2}\partial_{\beta}h_{00}=-\frac{c^{2}}{4}%
\partial_{\beta}\left( \overline{h}_{00}+\overline{h}_{\alpha\alpha}\right) .
\label{forcestat}
\end{equation}

From Eqs. (\ref{Rlin}) and (\ref{gauge}) we can write the equations for $
\overline{h}_{00}$ and $\overline{h}_{\alpha \beta }$ as
\begin{eqnarray}
\partial _{\gamma \gamma }\overline{h}_{00} &=&H,  \label{dggh00} \\
\partial _{\gamma \gamma }\overline{h}_{\alpha \beta } &=&2\partial _{\alpha
\beta }\sigma +\partial _{\alpha \gamma }\overline{h}_{\beta \gamma
}+\partial _{\beta \gamma }\overline{h}_{\alpha \gamma }-\delta _{\alpha
\beta }H,  \label{dgghab}
\end{eqnarray}%
where $\delta _{\alpha \beta }$ is Kroenecker delta, and 
\begin{equation}
H\equiv \partial _{ik}\overline{h}^{ik}=\partial _{\gamma }\left( \frac{
\partial _{\alpha \beta }\sigma \overline{h}_{\alpha \beta }}{3\partial
_{\delta }\sigma \partial _{\delta }\sigma }\partial _{\gamma }\sigma
\right) .  \label{heq}
\end{equation}%
In particular, one has $\partial _{\gamma \gamma }\overline{h}_{\alpha
\alpha }=-H$, which indicates that the effective potential $c^{2}\left( 
\overline{h}_{00}+\overline{h}_{\alpha \alpha }\right) /4$ is harmonic. An
estimation of the expected force for a solution of the type (\ref{dsigstat})
is%
\begin{equation}
\left\vert f_{\beta }\right\vert \sim \sqrt{8\pi G_{0}\varepsilon _{0}\left(
d\lambda /d\omega \right) _{\phi _{0}}}E.  \label{fstatest}
\end{equation}

Another simple case where the particular solutions considered can exist is
for propagating electromagnetic fields in vacuum, not dependent on a
Cartesian coordinate, which we denote as $z$. In effect, solenoidal electric
and magnetic fields can be represented in that case as ($\mathbf{e}_{z}$ is
the unit vector along the $z$ direction)%
\[
\mathbf{E}\left( x,y,t\right) =\mathbf{\nabla }\Phi \times \mathbf{e}_{z},\;%
\mathbf{B}\left( x,y,t\right) =\Psi \left( x,y,t\right) \mathbf{e}_{z}, 
\]%
while Faraday and Amp\`{e}re-Maxwell equations require that%
\[
\Psi \left( x,y,t\right) =\frac{1}{c}\partial _{0}\Phi ,\;\eta ^{ik}\partial
_{ik}\Phi =0. 
\]

In this way, Eq. (\ref{philin}) is written as%
\[
\partial ^{l}\sigma \partial _{l}\sigma =\frac{8\pi G_{0}\varepsilon
_{0}\left( d\lambda /d\omega \right) _{\phi _{0}}}{c^{4}}\partial ^{l}\Phi
\partial _{l}\Phi , 
\]%
with the possible solutions, satisfying (\ref{dalamb}),%
\begin{equation}
\partial _{l}\sigma =\pm K\partial _{l}\Phi .  \label{desigvar}
\end{equation}%
Again, for $\left( d\lambda /d\omega \right) _{\phi _{0}}<0$ solutions exist
with the role of electric and magnetic fields interchanged.

The specific, non-electromagnetic force on a stationary object is given in
this case by%
\begin{equation}
f_{\beta}=-\frac{c^{2}}{4}\partial_{\beta}\left( \overline{h}_{00}+\overline{%
h}_{\alpha\alpha}\right) +c^{2}\partial_{0}\overline{h}_{0\beta}.
\label{forcevar}
\end{equation}

A direct operation with Eq. (\ref{Rlin}) allows to write%
\[
\eta^{ik}\partial_{ik}f_{\beta}=-c^{2}\partial_{00}\left[ \partial_{\beta
}h+\partial_{\beta}\sigma\left( 1+\frac{\partial_{lm}\sigma\overline{h}^{lm}%
}{\partial^{l}\sigma\partial_{l}\sigma}\right) \right] , 
\]
from which, using (\ref{desigvar}), a simple estimation gives a result
similar in form to (\ref{fstatest}).

Finally, we consider a case analogous to the previous one, where now the
free electromagnetic field has rotational symmetry around the $z$ direction.
Using spherical coordinates in three-dimensional space we can write in this
case, for the solenoidal electric and magnetic fields ($\mathbf{e}_{\varphi
} $ is the unit vector in the local azimuthal direction) 
\[
\mathbf{E}\left( r,\theta ,t\right) =\frac{1}{r\sin \theta }\mathbf{\nabla }%
\Phi \times \mathbf{e}_{\varphi },\;\mathbf{B}\left( r,\theta ,t\right)
=\Psi \left( r,\theta ,t\right) \mathbf{e}_{\varphi }, 
\]%
and now the Faraday and Amp\`{e}re-Maxwell equations require that%
\begin{eqnarray}
\Psi \left( r,\theta ,t\right) &=&\frac{1}{cr\sin \theta }\partial _{0}\Phi ,
\nonumber \\
\partial _{00}\Phi &=&\frac{\partial ^{2}\Phi }{\partial r^{2}}+\frac{\sin
\theta }{r^{2}}\frac{\partial }{\partial \theta }\left( \frac{1}{\sin \theta 
}\frac{\partial \Phi }{\partial \theta }\right) .  \label{phirtheta}
\end{eqnarray}%
It is important to note that $\Phi $ does not satisfy a D'Alembert equation
like the one to be satisfied by $\sigma $ (Eq.(\ref{dalamb})).

Eq. (\ref{philin}) is now written as
\begin{equation}
\left( \partial _{0}\sigma \right) ^{2}-\left\vert \mathbf{\nabla }\sigma
\right\vert ^{2}=\frac{8\pi G_{0}\varepsilon _{0}\left( d\lambda /d\omega
\right) _{\phi _{0}}}{c^{4}r^{2}\sin ^{2}\theta }\left[ \left( \partial
_{0}\Phi \right) ^{2}-\left\vert \mathbf{\nabla }\Phi \right\vert ^{2}\right]
.  \label{philrtheta}
\end{equation}%
Special solutions of the kind considered can be obtained noting that Eq. (
\ref{philrtheta}) is satisfied if the following equations hold%
\begin{eqnarray*}
\mathbf{n}\partial _{0}\sigma -\mathbf{\nabla }\sigma  &=&\frac{\gamma K}{r\sin \theta }\left( \mathbf{m}\partial _{0}\Phi -\mathbf{\nabla }\Phi
\right) , \\
\mathbf{n}\partial _{0}\sigma +\mathbf{\nabla }\sigma  &=&\frac{K}{\gamma r\sin \theta }\left( \mathbf{m}\partial _{0}\Phi +\mathbf{\nabla }\Phi
\right) ,
\end{eqnarray*}%
where $\gamma $ is an arbitrary function of ($r,\theta ,t$), and $\mathbf{n}$
and $\mathbf{m}$ are two arbitrary unit vectors. Solving these equations for 
$\partial _{0}\sigma $ and $\nabla \sigma $ one hass%
\begin{eqnarray}
\partial _{0}\sigma  &=&\frac{K}{2\gamma r\sin \theta }\left[ \left(
1+\gamma ^{2}\right) \mathbf{m}\cdot \mathbf{n}\partial _{0}\Phi +\left(
1-\gamma ^{2}\right) \mathbf{n}\cdot \mathbf{\nabla }\Phi \right] ,
\label{d0sigma} \\
\mathbf{\nabla }\sigma  &=&\frac{K}{2\gamma r\sin \theta }\left[ \left(
1-\gamma ^{2}\right) \mathbf{m}\partial _{0}\Phi +\left( 1+\gamma
^{2}\right) \mathbf{\nabla }\Phi \right] .  \label{gradsigma}
\end{eqnarray}

One can now write Eq. (\ref{dalamb}) as
\begin{equation}
\partial _{0}\left( \partial _{0}\sigma \right) =\mathbf{\nabla }\cdot
\left( \mathbf{\nabla }\sigma \right) ,  \label{eqgamma}
\end{equation}%
where $\partial _{0}\sigma $ and $\nabla \sigma $ are replaced by their
expressions (\ref{d0sigma}) and (\ref{gradsigma}). In this way, solved Eq. (
\ref{phirtheta}) for $\Phi $, and given arbitrary expressions for the unit
vectors $\mathbf{n}$ and $\mathbf{m}$, Eq. (\ref{eqgamma}) turns out to be
an equation for the function $\gamma $, that once solved allows to obtain
the function $\sigma $ using (\ref{d0sigma}) and (\ref{gradsigma}). The
arbitrariness of the units vectors employed only means that the system of
equations considered admits a large class of solutions. The actual solution
in a real situation must correspond to unit vectors determined by the
solution itself. Considering that the physical problem is determined by the
scalar $\Phi $ one could take 
\[
\mathbf{n}=\mathbf{m}=\frac{\mathbf{\nabla }\Phi }{\left\vert \mathbf{\nabla 
}\Phi \right\vert }. 
\]

\subsection{Conclusions}

It was shown that a Brans-Dicke type of theory, with interaction of
Bekenstein's type between the Maxwell and the scalar fields, admits a special
kind of solutions in which the metric effect of the electromagnetic field is
much larger than in general relativity. This kind of solution arises for
particular electromagnetic field distributions, most notably in all static
cases, and also for propagating fields with Cartesian or axial symmetry, in
the regions outside the electromagnetic sources. Of course, even if the
theory is realist, the actual realization of this kind of solutions depends
on he possibility of satisfying the necessary boundary conditions between
the regions outside and inside the sources. As the theory is considered as
an effective one, and at the classical level, it is not easy to
ascertain the correct boundary conditions with real, microscopic sources. A
related point to be noted is that, from Eq. (\ref{Rlin}), the "forcing" of
the metric depends on the second derivative of $\sigma$ or, through (\ref%
{dsigstat}), on the second derivative of the electrostatic potential. This
means that in the case of a homogeneous electric field, the actual solution
is determined solely by the boundary conditions. It is thus expected that
inhomogenous fields are preferred in order to seek solutions of the type
considered.

As for the magnitude of the expected forces, from expression (\ref{fstatest}
) one can estimate that for a field of about 10 $kV/m$ and $\sqrt{\left(
d\lambda /d\omega \right) _{\phi _{0}}}\sim 1$ the corresponding specific
force is about 1 $\mu N/Kg$. Although small and difficult to be separated
from the forces of electric origin that necessarily should appear in an
experiment, such non-electrical forces can be measured with present
technology, as is well documented in\cite{Larue1}, \cite{larue2}.
It is also remarkable that forces of the type and magnitude just discussed have been reported in inhomogeneous electrostatic fields\cite{datta_ming}.

With the above considerations it is concluded that experiments of relative
simplicity are possible to test the possibility of scalar-tensor theories
with Maxwell-scalar fields interaction.

\end{document}